\documentclass[twocolumn,prb,eqsecnum,showpacs,preprintnumbers,amsmath,amssymb]{revtex4}

\usepackage{graphicx}
\usepackage{dcolumn}
\usepackage{bm}
\newcommand{\gz}{\mathbb{Z}} 

\begin{document}
\title{ Magnetic scattering of Dirac fermions 
in topological insulators and graphene }
\author{Alex Zazunov, Arijit Kundu, Artur H\"utten, and Reinhold Egger}
\affiliation{Institut f\"ur Theoretische Physik,
Heinrich-Heine-Universit\"at, D-40225 D\"usseldorf, Germany}

\date{\today}
\begin{abstract}
We study quantum transport and scattering of massless Dirac fermions
by spatially localized static magnetic fields.  
The employed model describes in a unified manner 
the effects of orbital magnetic fields, Zeeman and exchange 
fields in topological insulators, and the
pseudo-magnetic fields caused by strain or defects in monolayer graphene.  
The general scattering theory is formulated, and for radially symmetric
fields, the scattering amplitude and the total and transport cross sections
are expressed in terms of phase shifts. As applications, we 
study ring-shaped magnetic fields (including the 
Aharanov-Bohm geometry) and scattering by magnetic dipoles.
\end{abstract}


\maketitle
\section{Introduction}

The recent theoretical prediction and subsequential experimental verification
of the conducting surface state existing in a 
\textit{strong topological insulator} (TI)
has generated a burst of activity, reviewed in Refs.~\onlinecite{hasan} and
\onlinecite{qi}.   
In a TI, strong spin-orbit couplings and band inversion 
conspire to produce a unique time-reversal invariant topological state 
different from a conventional band insulator.
Using Bi$_2$Se$_3$ as reference TI material, one finds a
rather large bulk gap $\approx 0.3$~meV, and surface probe experiments 
have provided strong evidence for the topologically protected gapless surface 
state.\cite{xia} The measured spin texture of the surface 
is well described by two-dimensional (2D) massless Dirac fermions,
where the spinor wavefunction has precisely two entries 
corresponding to physical spin.  Under this ``relativistic'' description,
spin and momentum are always perpendicular, and the
surface state is stable against the effects of weak disorder and
weak interactions due to the underlying topological protection.\cite{hasan,qi}
Useful insights can then already be 
obtained from a noninteracting  disorder-free
description.  Massless 2D Dirac fermions are also realized in 
single carbon monolayers of \textit{graphene}, for 
reviews see Refs.~\onlinecite{review}, \onlinecite{castro} and 
\onlinecite{gusynin}.  
The limit of ballistic transport in this 2D material seems experimentally
within reach,\cite{andrei,bolotin} 
and graphene experiments have reported characteristic Dirac fermion
signatures, e.g., Klein tunneling.\cite{klein1,klein2} 
In most experiments performed so far, correlation effects turned out to 
be weak, and it is again of interest to examine the single-particle theory 
in the absence of disorder. 
In contrast to the TI surface case, the two entries of the
spinor wavefunction in graphene correspond to the two atoms in the 
basis of graphene's honeycomb lattice.
Furthermore, there are four Dirac fermion ``flavors''
in graphene, due to the physical spin and 
the $KK'$ (``valley'') orbital degeneracy.\cite{castro}  
Another condensed-matter realization of Dirac fermions
is given by the quasiparticles in $d$-wave superconductors,\cite{dwave}
but to be specific, 
we here focus on the surface state in a TI and on monolayer graphene.

The fact that the electronic properties of both these applications
correspond to massless 2D Dirac fermions
calls for a unified description of their transport properties.
In the graphene context, much theoretical effort has been devoted to 
advancing the scattering theory of Dirac fermions in electrostatic 
potentials,\cite{castro,novikov} in particular for the Coulomb 
impurity.\cite{coulomb} In this paper, we instead study
the scattering of massless Dirac fermions by
a local \textit{magnetostatic perturbation}.  
The model, see Eq.~\eqref{dirac} below,
describes, in a unified manner, the effects of 
spatially inhomogeneous orbital magnetic fields, 
exchange-mediated fields due to adjacent ferromagnetic (FM) layers and
Zeeman fields in topological insulators, as well as strain- or defect-induced  
pseudo-magnetic fields in graphene.  
For the Schr\"odinger fermions realized in 2D semiconductor electron gases, 
such perturbations, e.g.,
magnetically defined barriers, steps, and quantum wells, 
have been investigated both 
theoretically\cite{magbar1,magbar2,magbar3,magbar4} and 
experimentally.\cite{magbarexp,heinzel}
The desired magnetic field profiles were generated by deposition of
lithographically patterned FM layers on 
top of the sample. For soft FM materials,
one can change the magnetization orientation by weak magnetic fields.
Another possibility is to use a type-II superconductor film 
instead of the FM layer.\cite{sup1,sup2}

Previous theory work on TIs in inhomogeneous magnetic fields has
addressed only a few setups.  For the transmission of an electron through a  
magnetic barrier (assumed homogeneous in the transverse direction), 
as a function of either exchange field or applied bias voltage,
Mondal \textit{et al.}\cite{mondal}  predict 
an oscillatory behavior or even a complete suppression of
the transmission probability, and hence of the conductance.
A spin valve geometry with two adjacent magnetic barriers, 
characterized by non-collinear exchange fields, has also been
studied.\cite{nagaosa}   As a model for a classical magnetic impurity,
the spin-resolved density of states was calculated for 
a disc-shaped magnetic field profile.\cite{disc1,disc2}  

For graphene, a vector potential perturbation can again be due to 
external orbital fields, but may also describe the effects of
strain\cite{castro,strain1,strain2,strain3,strain4,strain5} 
and dislocations or other topological defects.\cite{voz} 
Several theoretical works have addressed 
aspects of the electronic structure and the
transmission properties for Dirac fermions in graphene 
in the presence of inhomogeneous magnetic fields.
The simplest case is encountered for effectively 1D problems with
translational invariance in the (say) $y$-direction, e.g., for a magnetic
step or a magnetic barrier.\cite{ademarti,ademarti2,masir1}
For suitable 1D magnetic field profiles, it is possible to have
magnetic waveguides (along the $y$-direction),\cite{lambert,ghosh} 
where electron-electron interaction effects play an important role.\cite{haus1}
Periodic magnetic fields, i.e., 1D magnetic superlattices, have also been 
addressed.\cite{tahir,luca,masir2,tan}
For radially symmetric fields, total angular momentum 
conservation again simplifies the problem and gives an effective 1D
theory.  This has  allowed for studies of quantum dot or antidot 
geometries,\cite{ademarti,ademarti2,kory,masir3} where true bound states, 
not affected by Klein tunneling, may exist.  In quantum dot setups,
interaction effects become important for strong confinement.\cite{haus2}
When the vector potential corresponds to an infinitely thin solenoid, 
we encounter an ultra-relativistic Dirac fermion generalization of the 
celebrated Aharonov-Bohm (AB) calculation.\cite{ab,flux} This generalization 
was discussed before,\cite{ruijsenaars,hagen1,hagen2,hagen3,giacconi,sakoda} 
and exact results for the transmission amplitude can be deduced.
Recent studies have also 
addressed the current induced by an AB flux\cite{jackiw}
and the behavior of the conductance when the chemical potential
is precisely at the Dirac neutrality point.\cite{katsnelson}
Such an AB conductance can be probed 
experimentally in ring-shaped graphene devices.\cite{ringexp1,ringexp2}

In this article, we formulate a general scattering theory approach
for massless 2D Dirac fermions in the presence of such magnetic
perturbations.  In Sec.~\ref{sec2}, we introduce the model 
and outline its application to graphene and topological insulators. 
In Sec.~\ref{sec3}, we formulate the general scattering theory, 
previously given for electrostatic potentials,\cite{novikov} 
for the magnetic case.  The scattering amplitude and cross section are 
specified, and we discuss the Born approximation in Sec.~\ref{sec3a}. For the
radially symmetric case, total angular momentum conservation allows
to express the scattering amplitude in terms of phase shifts for
given total angular momentum, see Sec.~\ref{sec3b}.
In sections~\ref{sec4} and \ref{sec5}, we present applications of this
formalism.  First, in Sec.~\ref{sec4}, we discuss scattering by 
a \textit{magnetic dipole} within the Born approximation.  Second, 
in Sec.~\ref{sec5}, we consider \textit{ring-shaped field profiles}.  
This case also contains the AB solenoid in a certain limit, 
see Sec.~\ref{sec5b}, and we discuss how our phase-shift analysis 
recovers known results for the AB effect.  
We also address the scattering resonances appearing
due to quasi-bound states in such a ring-shaped
magnetic confinement.  Finally, some concluding remarks 
can be found in Sec.~\ref{sec6}.

\section{Dirac fermions in graphene and topological insulators} \label{sec2}

\subsection{Model}\label{sec2a}

In this section, we describe the 2D Dirac fermion model 
for electronic transport in graphene or the TI surface studied in this work.
As perturbations, we first allow for an external static 
vector potential, ${\bm A}({\bm r})=(A_x,A_y,A_z)$ with 
${\bm r}=(x,y)$, which is included
by minimal coupling and describes orbital magnetic fields 
and (for graphene) strain-induced pseudo-magnetic fields. 
In fact, for those applications we have $A_z=0$, see below.
In addition, for the TI case, we allow for a Zeeman field 
or for exchange fields caused by nearby ferromagnets, whose 
components are contained in the field ${\bm M} ({\bm r})=(M_x,M_y,M_z)$, 
where prefactors such as the Bohr magneton or the Land{\'e} factor
are included.
With the Pauli matrices ${\bm \sigma}=(\sigma_x,\sigma_y,\sigma_z)$
and the momentum operator ${\bm p}=-i\hbar (\partial_x,\partial_y,0)$,
the single-particle model reads ($e>0$)
\begin{equation}\label{dirac1}
H =  v_F{\bm \sigma} \cdot \left({\bm p} + \frac{e}{c} {\bm A}({\bm r})\right)
+ {\bm \sigma}\cdot {\bm M}({\bm r}) - eV({\bm r}).
\end{equation}
The Fermi velocity in graphene is $v_F\simeq 10^6$~m/s,
while for a TI surface state, a typical value\cite{xia} for Bi$_2$Se$_3$ is 
$v_F \approx 5\times 10^5$~m/s. 
The low-energy Hamiltonian in Eq.~\eqref{dirac1} 
is valid on energy scales 
close to the neutrality level (Dirac point), well below the 
bulk band gap for the TI case and within a window of size $\approx 0.5$~eV 
for graphene.  
Both the vector potential ${\bm A}$ and the field ${\bm M}$ can 
now be combined to a vector field
\begin{equation}\label{Lambdadef}
{\bm \Lambda} ({\bm r}) \equiv {\bm A}+ \frac{c}{e v_F} {\bm M},
\end{equation}
which contains all considered ``magnetic'' perturbations.
Interesting physics also follows in the presence of both 
${\bm \Lambda}({\bm r})$ and a scalar potential $V({\bm r})$, but 
we put $V({\bm r})=0$ below.

For the formulation of the scattering theory, it is convenient to 
employ cylindrical coordinates, $x=r\cos\phi$ and $y=r\sin\phi$, with
unit vectors $\hat e_r = (\cos \phi, \sin \phi,0),$
 $\hat e_\phi = (-\sin \phi, \cos \phi,0)$, and $\hat e_z$.
With ${\bm \Lambda}= \Lambda_r \hat e_r+\Lambda_\phi \hat e_\phi+
\Lambda_z \hat e_z,$ Eq.~\eqref{dirac1} takes the compact form
\begin{eqnarray}\label{dirac}
H &=& v_F e^{-i\phi\sigma_z/2} \tilde H e^{i\phi\sigma_z/2},\\ \nonumber
\tilde H &=& \left(-i\hbar \partial_r+\frac{e}{c} \Lambda_r\right)
\sigma_x + \left ( \frac{1}{r} J_z + \frac{e}{c} 
\Lambda_\phi\right) \sigma_y ,
\end{eqnarray} 
where the total angular momentum operator is
\begin{equation}\label{jdef}
J_z = -i \hbar \partial_\phi + \hbar \sigma_z/2.
\end{equation} 
For the case of azimuthal symmetry, $\partial_\phi\Lambda_{\phi,r}=0$,
this is a conserved quantity, $[J_z,H]=0$, with
eigenvalues $\hbar j$ for half-integer $j$.
In Eq.~\eqref{dirac} we have put $\Lambda_z=0$, which is the case for all
fields studied below.

Let us now discuss how Eq.~(\ref{dirac}) relates to the surface
states in a \textit{strong topological insulator}, where
the spin direction is tangential to the surface and perpendicular to momentum. 
For a given two-component spinor wavefunction $\psi({\bm r})$, 
the spin (${\bm s}$) and particle
current (${\bm j}$) density operators are\cite{hasan,qi} 
\begin{eqnarray}\label{spinpauli}
{\bm s}({\bm r}) &=& \frac{\hbar}{2}\psi^\dagger \left( \hat e_z \times 
{\bm \sigma} \right) \psi ,\\ \nonumber
{\bm j}({\bm r}) &=& v_F\psi^\dagger  
\left(\sigma_x\hat e_x+\sigma_y\hat e_y\right) \psi,
\end{eqnarray}
i.e., both spin and current are confined to the surface and obey 
${\bm s} \cdot {\bm j} = 0$.  Under stationary conditions,
the continuity equation,
$\partial_t  (\psi^\dagger \psi ) + \sum_{i=x,y} \partial_i j_i = 0$,
implies the relation
\begin{equation} \label{conteq}
\sum_{i=x,y} \partial_{i} \left( \psi^\dagger \sigma_i \psi \right)= 0,
\end{equation}
which is linked to the unitarity property of the scattering
matrix.  Equation (\ref{dirac}) then allows to describe the 
following setups for the TI
surface state.  First, an orbital magnetic field has only effects when
it is oriented perpendicular to the surface, ${\bm B}_{\rm orb}= 
B_z(r,\phi) \hat e_z$. In cylindrical coordinates,  we 
can then choose some gauge for the vector potential ${\bm A}$ such that
\begin{equation}\label{orbitalfield}
B_z(r,\phi) = \frac{1}{r} \left( \partial_r(rA_\phi) -
\partial_\phi A_r \right),
\end{equation}
while $A_z$ drops out and is put to zero.
Second, to describe the coupling of surface Dirac fermions 
to an \textit{in-plane exchange field} ${\bm H}({\bm r})= (H_x, H_y, 0)$,
e.g., due to the magnetization of a nearby FM layer,
we write ${\bm \Lambda}=(c/ev_F){\bm M}$ with ${\bm M}=(-H_y,H_x,0)$,  
where we used the spin Pauli matrices in Eq.~(\ref{spinpauli}).
For a Zeeman field, we can proceed in complete analogy where ${\bm H}$
now denotes the Zeeman field.  A Zeeman or exchange field oriented
along the $\hat e_z$ direction can open a gap in the spectrum, and
here we assume that such fields are not present.  While the orbital 
field breaks time reversal invariance,  the Zeeman or exchange fields 
represent a time-reversal invariant perturbation. 
Then ${\bm \Lambda}$ is determined by the 
magnetic field itself, and hence is not a gauge field anymore.

Next we turn to \textit{graphene},  where the Pauli matrices
${\bm \sigma}$ are related to the two triangular sublattices constituting
graphene's honeycomb lattice.  We assume that no spin-flip mechanisms 
are relevant, i.e., physical spin is conserved. 
We can then focus on one specific Dirac fermion flavor
with fixed valley index and spin direction. 
This excludes exchange or Zeeman fields, i.e., 
we put ${\bm M}=0$ and hence ${\bm \Lambda}={\bm A}$ for graphene. 
Note that Zeeman fields in graphene are generally small compared to 
orbital fields.\cite{haus1}   
Moreover, we consider only smoothly varying vector potentials such that it is
indeed sufficient to retain only one $K$ point.\cite{ademarti} 
Equation \eqref{dirac} can then describe the following cases.  
First, we may have an orbital magnetic field, precisely as for the TI case.
Second, pseudo-magnetic fields generated by strain-induced 
forces,\cite{castro,strain1,strain2,strain3,strain4,strain5}
or by various types of defects, e.g., dislocations,\cite{voz}
also correspond to a vector potential, where
time reversal invariance implies that ${\bm A}$ has 
opposite sign at the two $K$ points. 
${\bm A}({\bm r})$ can then be expressed explicitly in terms
of the strain tensor,\cite{voz} where the resulting
pseudo-magnetic field is also oriented along the $\hat e_z$ axis and
$A_z=0$.  In addition, strain  causes a scalar potential $V({\bm r})$, which 
is, however, strongly reduced by screening effects.
The combination of orbital and pseudo-magnetic fields
may allow to design a valley filter, since the total (orbital plus
pseudo-magnetic) fields can differ significantly at both 
$K$ points.\cite{valleyfilter}  

\subsection{Multipole expansion}\label{sec2b}

Our scattering theory approach considers magnetic fields [described by
${\bm \Lambda}({\bm r})$ in Eq.~\eqref{dirac}] that smoothly vary on the 
scale of a lattice spacing and constitute a \textit{local}\ perturbation, i.e.,
a well-defined cylindrical multipole expansion exists.  Furthermore, 
$\Lambda_z=\Lambda_r=0$ is assumed throughout.  As we show below, 
for orbital fields we can choose a gauge where $A_r=0$.  For strain-induced 
fields, strictly speaking, the problem is not gauge invariant, and we cannot
impose gauge conditions.  However, in a more narrow sense, a gauge degree of 
freedom still exists.\cite{voz}
For $r\to \infty$, with complex-valued coefficients 
$\alpha^{(\phi)}_{l,m}= \left( \alpha^{(\phi)}_{l,-m}\right)^*$, we
then have the multipole expansion
\begin{equation}\label{decomp}
\Lambda_{\phi}(r,\phi) =  \frac{\alpha\Phi_0}{2\pi r} +
\sum_{l=2}^\infty \sum_{m=-\infty}^\infty
 \frac{e^{im\phi}}{r^l} \alpha^{(\phi)}_{l,m},
\end{equation}
where $\alpha$ denotes the total flux in units of the flux
quantum $\Phi_0=2\pi\hbar c/e$.

Let us now address the orbital magnetic field case,
${\bm \Lambda}={\bm A}$, where we can exploit 
gauge invariance. We start from a more general situation with $A_r\ne 0$,
expressed as in Eq.~\eqref{decomp} with coefficients $\alpha_{l,m}^{(r)}$,
and also allow for nonzero coefficients $\alpha_{l=1,m\ne 0}^{(\phi)}$.
We now show that one 
can choose a gauge where $A_r=0$ and $\alpha_{1,m\ne 0}^{(\phi)}=0$.
Indeed, gauge invariance implies that for arbitrary functions $g(x,y)$,
we are free to replace $A_i\to A_i+\partial_i g$.
Using a multipole expansion for $rg(r,\phi)$ with coefficients $g_{l,m}$,
an equivalent gauge choice thus follows by the replacement
\begin{eqnarray*}
\alpha_{l,m}^{(\phi)} &\to & \alpha_{l,m}^{(\phi)}+ im g_{l,m} ,
\\ \alpha_{l,m}^{(r)} &\to &\alpha_{l,m}^{(r)}-(l-1) g_{l,m} .
\end{eqnarray*}
We then choose the gauge function
\[
g_{l>1,m}= \frac{\alpha_{l,m}^{(r)}}{l-1},\quad
g_{l=1,m\ne 0}=\frac{i\alpha_{1,m}^{(\phi)}}{ m}.
\]
In the new gauge, we arrive at Eq.~(\ref{decomp}) plus the radial component
\[
A_r = \sum_{m}  \frac{e^{im\phi}}{r} \alpha^{(r)}_{1,m} .
\]
Using Eq.~\eqref{orbitalfield}, the orbital field expansion (with $r>0$) reads 
\[
B_z(r,\phi) = -\sum_{l=1}^\infty \sum_{m=-\infty}^\infty
\frac{e^{im\phi}}{ r^{l+1}} \left[ (l-1) \alpha^{(\phi)}_{l,m} 
+ im\delta_{l,1} \alpha_{1,m}^{(r)} \right] . 
\]
The $m=0$ term in $A_r$ neither generates flux nor magnetic fields
and can be omitted.
Magnetic field profiles with $\alpha_{1,m}^{(r)}\ne 0$ arise only
in time-dependent settings and will not be studied here. As
a consequence, the radial component vanishes, $A_r=0$, and we arrive
at Eq.~\eqref{decomp}.

\section{Scattering theory} \label{sec3}

For given energy $E =\hbar v_F k$, where $k> 0$ throughout,
the Dirac equation, $H\psi=E\psi$ with Eq.~\eqref{dirac}, 
has scattering solutions that we wish to obtain in the presence
of magnetic perturbations of the type in Eq.~\eqref{decomp}.
The solution for $E=-\hbar v_F k$ 
follows simply by reversing the sign of the lower 
spinor component.\cite{gusynin}  We are then looking for a solution 
$\psi(r,\phi)=\psi_{\rm in}+\psi_{\rm out}$ consisting, in 
the asymptotic regime $r\to \infty$, of a plane wave ($\propto e^{ikx}$)
propagating along the positive $x$-direction,
\begin{equation}\label{incoming} 
\psi_{\rm in}(r,\phi) = \frac{1}{\sqrt{2}} e^{ikr\cos\phi}
\left(\begin{array}{c} 1\\ 1\end{array}\right),
\end{equation} 
plus the scattered outgoing spherical wave,\cite{newton} 
\begin{equation}\label{outgoing}
\psi_{\rm out}(r,\phi) = F(\phi) \frac{e^{ikr}}{\sqrt{-2ir}} 
\left(\begin{array}{c} 1\\ e^{i\phi}\end{array}\right).
\end{equation}   
We adopt the same normalization conventions as Novikov.\cite{novikov}
{}From Eq.~(\ref{spinpauli}) we see that
the incoming current density is ${\bm j}_{\rm in}=v_F \hat e_x$  
while (for the TI case) the spin density is $(\hbar/2)\hat e_y$.
Equation \eqref{outgoing}  defines the 
\textit{scattering amplitude} $F(\phi)$ for an outgoing wave
deflected under the scattering angle $\phi$.
The resulting scattered current density implies the standard definitions of
the differential ($d\sigma/d\phi$), total ($\sigma_{\rm tot}$) 
and transport  ($\sigma_{\rm tr}$) scattering cross 
sections,\cite{novikov,newton} respectively:
\begin{eqnarray}\label{cross}
\frac{d \sigma}{ d \phi }  &=& |F(\phi)|^2,\\ \nonumber
\sigma_{\rm tot} &=&  \int_0^{2 \pi} d \phi \ |F(\phi)|^2 = 
\sqrt{\frac{8\pi}{k}} \ {\rm Im} F(0),\\ \nonumber
\sigma_{\rm tr} &=& \int_0^{2\pi} d\phi \ (1-\cos\phi) \ |F(\phi)|^2.
\end{eqnarray}
The second equality for $\sigma_{\rm tot}$ expresses the 2D optical theorem.
When a random distribution of magnetic perturbations is present, 
the inverse mean free path determining the conductivity is proportional to the
transport cross section.\cite{abrikosov}

\subsection{Born approximation}\label{sec3a}

For small perturbation $\Lambda_\phi(r,\phi)$, one can evaluate
the scattering amplitude within the first Born approximation.\cite{newton}  
Strictly speaking, the long-ranged part 
$\Lambda_\phi\propto \alpha/r$ in Eq.~\eqref{decomp} 
can not be treated perturbatively, and in this section we 
assume $\alpha=0$.  

The unperturbed state is the incoming plane wave $\psi_{\rm in}$, 
Eq.~\eqref{incoming}. Within lowest-order perturbation theory, 
the scattered wave obeys  
\begin{equation}\label{start}
[ H_0- E ] \psi_{\rm out} = -\frac{e v_F}{c} \Lambda_\phi \ \hat e_\phi \cdot 
{\bm \sigma}  \ \psi_{\rm in},
\end{equation}
where $H_0$ is the unperturbed Dirac Hamiltonian.
Multiplying both sides of Eq.~\eqref{start} by $H_0+E$ and noting that
in real-space representation, the retarded Green's function 
$(H_0^2-E^2)^{-1}$ is given by the Hankel
 function $H_0^{(1)}$,\cite{novikov,gradsteyn}
\begin{eqnarray*}
\psi_{\rm out}({\bm r}) & = & \frac{-i\pi}{2\sqrt{2} \hbar \Phi_0}
\int d^2{\bm r}' \ H_0^{(1)}(k|{\bm r}-{\bm r'}|) 
( {\bm \sigma}\cdot {\bm p}' + 
\hbar k) \\ &\times& 
\Lambda_\phi(r',\phi') \ [\hat e_{\phi'}\cdot{\bm \sigma}] \
e^{ikr'\cos\phi'} \left( \begin{array}{c} 1 \\ 1 \end{array}\right).
\end{eqnarray*}
The asymptotic large-$\rho$ behavior of the Hankel function
(where $\eta=1,2=\pm$) is\cite{gradsteyn}
\begin{equation}\label{asymp}
H^{(\eta)}_\nu(\rho) \simeq \sqrt{\frac{2 }{ \pi \rho }} \
e^{\pm i ( \rho - (2\nu+1)\pi/4)},
\end{equation}
which implies that $\psi_{\rm out}$ for $r\to \infty$ indeed
has the form in Eq.~(\ref{outgoing}). 
After some algebra, we obtain the scattering amplitude 
in Born approximation,
\begin{eqnarray} \nonumber
F(\phi)  &=& \frac{\sqrt{2\pi k}}{\Phi_0} e^{-i\phi/2}
\int_0^\infty r dr \int_0^{2\pi} d\phi' \ \sin\phi'
 \\ &\times& e^{-2ikr|\sin(\phi/2)|\sin\phi'} \
\Lambda_\phi(r,\phi'+\phi/2) .
\label{bornscatt}
\end{eqnarray}
For radially symmetric perturbations, $\partial_\phi\Lambda_\phi=0$,
the $\phi'$-integration can be done, and we obtain
\begin{eqnarray}\label{born2}
F(\phi) &=& -2\pi i \frac{\sqrt{2\pi k}}{\Phi_0}
e^{-i\phi/2}  \\ &\times& \nonumber
\int_0^\infty r dr \ J_1\left(2kr|\sin(\phi/2)|\right) \ \Lambda_\phi(r) ,
\end{eqnarray}
with the $J_1$ Bessel function.

\subsection{Radially symmetric case} \label{sec3b}

Next, we address the full (beyond Born approximation)
scattering solution for radially symmetric 
perturbations, ${\bm\Lambda}=\Lambda_\phi (r) \hat e_\phi$. 
In that case, the total angular momentum operator $J_z$ in Eq.~(\ref{jdef})
is conserved and has eigenvalues $\hbar j$ with $j\equiv m+1/2$ ($m\in \gz$).
We thus expand the spinor wavefunction in terms of angular momentum partial
waves $\psi_m(r) \equiv (f_m,ig_m)^T,$ 
\begin{equation}\label{expand2}
\psi(r,\phi) = e^{-i \phi \sigma_z / 2} \sum_{m=-\infty}^\infty
e^{i (m + 1/2) \phi} \  \psi_m (r) ,
\end{equation}
where the Dirac equation yields
\begin{equation}
\left [ -i\left( \partial_r + \frac{1} {2 r} \right) \sigma_x +
 \frac{m+1/2+\varphi(r)}{r}\sigma_y \right] \psi_m = k \psi_m.
\end{equation}
The magnetic flux (in units of the flux quantum $\Phi_0$)
enclosed by a circle of radius $r$ around the origin is
\begin{equation} \label{lll}
 \varphi(r) \equiv \frac{2\pi r}{\Phi_0} \Lambda_\phi(r) ,
\end{equation}
where $\alpha=\varphi(\infty)$ in Eq.~\eqref{decomp}. 
 The continuity relation (\ref{conteq}) must 
hold for each partial wave $\psi_m$ separately, and implies
\begin{equation}   \label{conteq2}
\partial_r \left( r \psi^\dagger_m \sigma_{x} \psi^{}_m \right)= 0 .
\end{equation}
Introducing dimensionless radial coordinates, $\rho \equiv k r$,
a closed equation for the upper component, $f_m(\rho)$, follows, 
\begin{eqnarray} \nonumber
&& \left[ \frac{1}{\rho} \partial_\rho ( \rho\partial_\rho ) + 1 -
\left( \frac{1}{4\rho^2} + W_m^2+ W'_m \right) \right] f_m = 0,\\ \label{fmeq}
&& \quad W_m(\rho) \equiv \frac{m+1/2+\varphi(\rho/k)}{\rho}  ,
\end{eqnarray}
where $W'_m\equiv \partial_\rho W_m$. The lower component is obtained from 
\begin{equation} \label{gmeq}
g_m (\rho) = -\left( \partial_\rho+ \frac{1}{2\rho} - W_m \right) f_m .
\end{equation}
These relations imply a general expression for the 
scattering amplitude $F(\phi)$
 under radially symmetric magnetic perturbations,
and thus for the various cross sections in Eq.~\eqref{cross}.
For $\rho\to\infty$, the term $\propto \alpha/r$ 
in Eq.~\eqref{decomp} dominates and
the general solution to Eq.~(\ref{fmeq}) is
given in terms of Hankel functions,
\begin{equation}\label{fmb}
f_m (\rho) = a_m H_{m+\alpha}^{(1)}(\rho) + b_m H_{m+\alpha}^{(2)}(\rho),
\end{equation}
with complex coefficients $a_m$ and $b_m$.
The lower spinor component then follows from Eq.~\eqref{gmeq},
\begin{equation}\label{gmb}
g_m (\rho) =  a_m H_{m+\alpha + 1}^{(1)} (\rho) + 
 b_m H_{m+\alpha+1}^{(2)}(\rho)  .
\end{equation}
The continuity relation (\ref{conteq2}) implies  
$a_m=b_m  e^{2 i \tilde \delta_m}$, i.e., the outgoing wave can differ 
from a free spherical wave only by a phase shift $\tilde\delta_m$,
which depends on the magnetic perturbation and is
determined in Sec.~\ref{sec5}.  Using the Bessel function expansion formula
\[
e^{i \rho \cos \phi} = \sum_{m\in\gz} i^m  e^{i m  \phi}  J_m(\rho)
\]
and the asymptotic behavior of $H_\nu^{(1,2)}$, see Eq.~\eqref{asymp}, we find 
\begin{equation} \label{bm}
b_m =\frac{ i^m}{2}  e^{-i\pi\alpha/2} .
\end{equation}
We then obtain the scattering amplitude in terms of phase shifts
as for the electrostatic case,\cite{novikov}
\begin{equation}\label{central}
F(\phi) = \frac{-i}{\sqrt{2 \pi k}} \sum_{m\in\gz}  
\left( e^{2 i \delta_m} - 1 \right) e^{im\phi} ,
\end{equation}
but $\delta_m$ includes the total flux $\alpha$,
\begin{equation}\label{phaseshift}
\delta_m \equiv \tilde \delta_m - \pi \alpha /2.
\end{equation}
As a consequence, qualitatively different effects beyond
the electrostatic case arise, such as the AB effect.
  The cross sections in Eq.~\eqref{cross} are then given by
\begin{eqnarray}\label{cross1}
\sigma_{\rm tot}&=& \frac{4}{ k} \sum_m \sin^2 \left(\delta_m\right), \\
\nonumber
\sigma_{\rm tr}&=& \frac{2}{k} \sum_m \sin^2 \left(\delta_{m+1}-
\delta_m\right). 
\end{eqnarray}
Scattering theory has thus been reduced to the determination
of the phase shifts $\delta_m$.
In the electrostatic case,\cite{novikov} the phase
shifts obey the symmetry relation $\delta_{m}=\delta_{-m-1}$, implying 
the absence of backscattering, $F(\pi)=0$.  
In the magnetic case under consideration here, in general this
symmetry relation breaks down, and hence backscattering
is not suppressed anymore, $F(\pi)\ne 0$.  This is closely
related to the fact that magnetic fields can confine
massless Dirac particles.\cite{ademarti}

\section{Magnetic dipoles}\label{sec4}

As a first application, we analyze the scattering of 
2D massless Dirac fermions by a fixed
magnetic dipole moment ${\bm m}$ located at position ${\bm r}=(0,0,h)$,
i.e., at a height $h$ above the origin of the 2D plane.
In that case no total flux is generated, $\alpha=0$.  
The results of this section are obtained under  the 
Born approximation, see Sec.~\ref{sec3a}.

\subsection{Perpendicular orientation}

\begin{figure}
\includegraphics[width=0.45\textwidth]{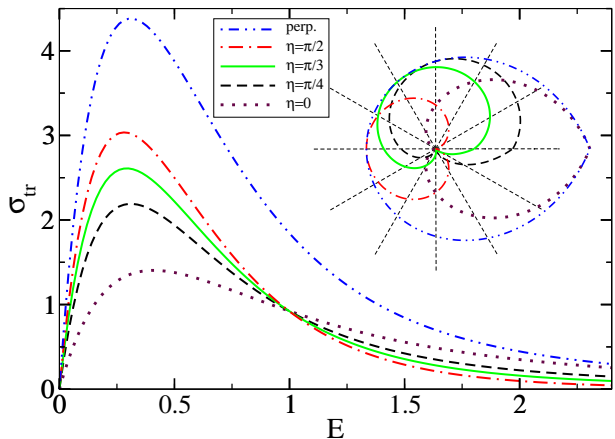}
\caption{\label{fig1}  (Color online) 
Born approximation results for the transport cross section $\sigma_{\rm tr}$
for scattering of massless Dirac fermions on a magnetic dipole. We show
$\sigma_{\rm tr}$ in units
of  $(2\pi)^4 (|{\bm m}|/\Phi_0)^2$ vs energy $E$ in units of $\hbar v_F/h$,
where $h$ is the distance of the dipole from the layer.  
The dot-dashed blue curve is for the perpendicular orientation, see
Eq.~\eqref{total0}, the other curves are for the parallel orientation
and several angles $\eta$ (cf.~legend), see Eq.~\eqref{total1}.
Upper right part:  Polar plot of the differential
cross section, $|F(\phi)|^2$, for $E=0.2$. 
Different curves correspond to the ones in the main panel.
}
\end{figure}

First, we consider a dipole moment oriented perpendicular to the
layer, ${\bm m}_\perp=m_\perp \hat e_z$, where we have the isotropic
(vector potential) perturbation 
\begin{equation}\label{mperp}
\Lambda_{\phi}(r) =m_\perp \frac{r}{(r^2+h^2)^{3/2}}.
\end{equation}
The Born approximation, see Eq.~\eqref{born2}, 
yields the scattering amplitude 
\begin{equation}
F_\perp(\phi)=-i e^{-i\phi/2} \sqrt{(2\pi)^3 k} \ (m_\perp/\Phi_0) 
 e^{-2kh\left|\sin(\phi/2)\right|},
\end{equation}
and the transport cross section is
\begin{equation}\label{total0}
\sigma_{{\rm tr},\perp} = (2\pi)^4 k \ (m_\perp/\Phi_0)^2  \tilde F_2(kh).
\end{equation}
Here we  define the functions
\begin{equation}\label{fnf}
\tilde F_{n}(x)=\frac{4}{\pi}
\int_{0}^{1} dt \ \frac{t^n e^{-4tx}}{\sqrt{1-t^2}},
\end{equation}
which can be expressed in terms of hypergeometric functions. 
As shown in Fig.~\ref{fig1}, after reaching a maximum around $kh\simeq 0.31$,
the transport cross section \eqref{total0} decreases
with increasing energy and approaches zero for $E\to \infty$.
Figure \ref{fig1} also shows a polar graph for the 
differential cross section, $d\sigma/d\phi=|F(\phi)|^2$,
at $kh=0.2$. Evidently, scattering by a dipole 
oriented along the $\hat e_z$ axis is almost isotropic.

\subsection{Parallel orientation}

If the dipole instead points parallel to the layer,
\begin{equation}
{\bm m}=m_\parallel  \left[ -\sin(\eta) \hat e_x+\cos(\eta)\hat e_y\right] ,
\end{equation}
where $\eta$ denotes an angle, we have
\begin{equation}
\Lambda_{\phi}(r,\phi)=\frac{m_\parallel}{h} 
\sin(\phi-\eta) \left( \frac{r^2}{(r^2+h^2)^{3/2}}-\frac{1}{r} \right),
\end{equation}
i.e., no radial symmetry is present.
Using Eq.~\eqref{bornscatt}, we now find the scattering amplitude
\begin{eqnarray} \label{11}
F_\parallel(\phi) &=&- e^{-i\phi/2} \sqrt{(2\pi)^3 k} \ (m_\parallel/\Phi_0)\\
&\times& \cos(\eta-\phi/2) e^{-2kh\left|\sin(\phi/2)\right|}, \nonumber
\end{eqnarray}
and thus the transport cross section 
\begin{equation} \label{total1}
\sigma_{{\rm tr},\parallel} = (2\pi)^4 k \ (m_\parallel/\Phi_0)^2 
\ \left[ \tilde F_4 + (\tilde F_2-2\tilde F_4)\cos^2(\eta) \right],
\end{equation}
where $\tilde F_n=\tilde F_n(kh)$.
This cross section depends on the orientation $\eta$ of the dipole
even for small energies, $kh\ll 1$. When averaging over $\eta$ (which
is equivalent to setting $\eta=\pi/4$), we find $\sigma_{{\rm tr},\parallel}=
(m_\parallel/m_\perp)^2 \sigma_{{\rm tr},\perp}/2$.
The transport cross section \eqref{total1} has a maximum
for an $\eta$-dependent energy, see Fig.~\ref{fig1}, and 
again approaches zero for $E\to \infty$.  
The differential cross section shown in Fig.~\ref{fig1} also reveals
a pronounced angular dependence tied to the orientation of the dipole.

\subsection{Bilayer graphene}

Let us briefly comment on the results under a quadratic
dispersion relation as realized in bilayer graphene.\cite{castro}
Repeating the Born approximation analysis,
the scattering amplitude is found to contain
an additional $\cos(\phi/2)$ factor modifying the above expressions.  
In fact, we find instead of Eqs.~\eqref{total0} and \eqref{total1}:
\begin{eqnarray}\nonumber
\sigma^{({\rm BLG})}_{{\rm tr},\perp} &=&  
(2\pi)^4 k \ (m_\perp/\Phi_0)^2  \ (\tilde F_2-\tilde F_4), \\ \label{blg}
\sigma^{({\rm BLG})}_{{\rm tr},\parallel} &=&
(2\pi)^4 k \  (m_\parallel/\Phi_0)^2 \\ \nonumber
&\times& \left [\tilde F_4-\tilde F_6+
(\tilde F_2-3 \tilde F_4+ 2 \tilde F_6) \cos^2(\eta) \right]  ,
\end{eqnarray}
with $\tilde F_n=\tilde F_n(kh)$.
The quoted expressions  hold for the quadratic dispersion relation of 
bilayer graphene,
and coincide with the results for the conventional Schr\"odinger case.
In contrast to the monolayer results for $\sigma_{{\rm tr},\parallel}$
in Eq.~\eqref{total1}, the transport cross section \eqref{blg} 
carries no $\eta$-dependence at low energies.  The latter is a distinctive
feature of 2D massless Dirac fermions.
Finally, we note that the $\sigma_{{\rm tr},\parallel}$ results
for $\eta=0$ in Fig.~\ref{fig1}
coincide with the bilayer result $\sigma_{{\rm tr},\perp}^{(\rm BLG)}$ (when
$m_\perp=m_\parallel$) for perpendicular orientation.

\section{Ring-shaped magnetic fields} \label{sec5}

\begin{figure}
\includegraphics[width=0.45\textwidth]{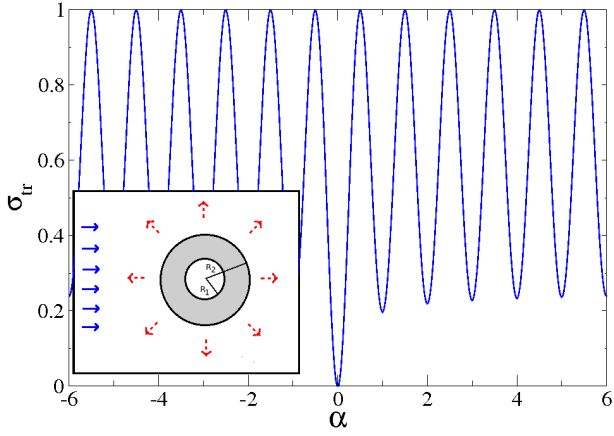}
\caption{\label{fig2} (Color online) 
Transport cross section $\sigma_{\rm tr}$ (in units of $2/k$) vs 
dimensionless flux $\alpha$ for a finite-width magnetic ring, see
Sec.~\ref{sec5c}, with $kR_1=0.01$ and $R_2=2R_1$.  (We here also
allow for $\alpha<0$.) 
The numerical results are close to the ideal AB prediction 
for the infinitely thin solenoid, $\sigma_{\rm tr}=(2/k)
\sin^2(\pi\alpha)$.  Inset:  Schematic scattering 
geometry.  The plane wave (blue solid arrows) coming in
along the $\hat e_x$ direction is scattered 
 by a ring-shaped magnetic field present 
for $R_1<r<R_2$ (shaded region). The outgoing 
spherical wave is indicated by red dashed arrows.
}
\end{figure}

In this section we consider the scattering states for 
a radially symmetric ring-shaped magnetic field.
The scattering setup is schematically sketched in the inset of Fig.~\ref{fig2}.

\subsection{Infinitesimally thin ring}
\label{sec5a}

Let us first study the exactly solvable model of 
an infinitesimally thin ring of radius $R$ around the origin,
where $\Lambda_\phi(r)$ follows from Eq.~\eqref{lll} with
\begin{equation}\label{aphi}
\varphi(r)= \alpha \Theta (r - R),
\end{equation}
where $\Theta$ is the Heaviside step function and, as before,
$\alpha$ is the dimensionless total flux through the ring surface area.
For the orbital field case, this implies
$B_z(r) = (\alpha \Phi_0  / 2 \pi R) \delta(r-R)$.
With $\rho=kr$ and ${\cal R}\equiv k R$, the solution to Eq.~(\ref{fmeq}) is
\begin{equation} \label{fmthin}
f_m (\rho)  = \left\{ \begin{array}{cc} a_m  J_m(\rho), & \rho<{\cal R}, \\
b_m  \left( e^{2 i \tilde \delta_m} H_{m+\alpha}^{(1)}(\rho) + 
H_{m+\alpha}^{(2)}(\rho) \right) , &   \rho>{\cal R} , \end{array} \right. 
\end{equation}
with $b_m$ in Eq.~(\ref{bm}).
The requirement of continuity of $\psi_m(r)$ at $r = R$,
together with Eq.~\eqref{gmeq}, 
leads to two boundary conditions for $f_m$. With ${\cal R}^\pm\equiv 
{\cal R}\pm 0^+$, they read
\begin{equation} \label{bc}
f_m \left({\cal R}^+\right) = f_m\left({\cal R}^-\right) ,\quad
f'_m\left ({\cal R}^+\right) - f'_m\left({\cal R}^-\right) =
\frac {\alpha}{{\cal R}} f_m({\cal R}) ,
\end{equation}
where again $f'=\partial_\rho f$.
The coefficient $a_m$ and the phase shift $\tilde \delta_m$ appearing
in Eq.~\eqref{fmthin} then
follow from the boundary conditions (\ref{bc}).
In particular, when $J_m({\cal R}) \neq 0$, 
the phase shift $\tilde \delta_m$ can be determined
by evaluation of the logarithmic derivative 
\begin{eqnarray} \label{calL}
{\cal L}_m & \equiv &  \frac{d \ln f_m(\rho={\cal R}^+)}{ d\rho} 
= \frac{\alpha}{{\cal R}} + \frac {J_m'({\cal R}) }{J_m({\cal R})} \\
\nonumber &=& \frac{m+\alpha}{{\cal R}} - 
\frac{J_{m+1}({\cal R})}{J_m({\cal R})},
\end{eqnarray}
where we used the second boundary condition in Eq.~(\ref{bc}).
As a result, with the Neumann function $Y_\nu$, we find
\begin{equation} \label{dmthin}
\tan \tilde \delta_m  = \frac{J'_{m + \alpha}({\cal R})- 
{\cal L}_m J_{m+\alpha}({\cal R})}
{Y'_{m + \alpha}({\cal R})- {\cal L}_m Y_{m+\alpha}({\cal R})},
\end{equation}
while $a_m$ is given by
\begin{equation} \label{amthin}
a_m = b_m  \frac{e^{2 i \tilde \delta_m} 
H_{m+\alpha}^{(1)}({\cal R}) + H_{m+\alpha}^{(2)}({\cal R})}{J_m({\cal R})} .
\end{equation}
Equation \eqref{dmthin} stays valid beyond the thin-ring limit
when a more general form for ${\cal L}_m$ is used, see Sec.~\ref{sec5c}.

For the special case $J_m({\cal R}) = 0$, Eq.~(\ref{fmthin}) implies
$e^{2 i \tilde \delta_m} = - H_{m+\alpha}^{(2)}({\cal R})/
H_{m+\alpha}^{(1)}({\cal R})$
and, using $f'_m ({\cal R}^+)=f'_m({\cal R}^-)$,
\[
a_m = b_m \frac{e^{2 i \tilde \delta_m}  \partial_{\cal R}
 H_{m+\alpha}^{(1)} ({\cal R}) + 
\partial_{\cal R} H_{m+\alpha}^{(2)}({\cal R})}{J'_m({\cal R})} .
\]
Equations~(\ref{dmthin}) and (\ref{amthin}) include
these relations when taking the limit
$J_m({\cal R}) \to 0$ and ${\cal L}_m \to \infty$.

\subsection{Aharonov-Bohm scattering amplitude}
\label{sec5b}

Let us first consider the $R\to 0$ limit of the above setting,
which corresponds to the pure solenoid case. This
allows us to study the Aharanov-Bohm (AB)
 effect for ultra-relativistic Dirac fermions.
In order to extract the singular part, we first rewrite $f_m(r)$
in Eq.~(\ref{fmb}) as
\begin{eqnarray}
f_m(r) &=& 2b_m \frac{e^{i \tilde \delta_m}}{ \sin (\pi\alpha )}
\Bigl[ \sin (\pi\alpha - \tilde \delta_m) \ J_{m+\alpha}(kr)
\nonumber \\ &+&  (-)^m \sin (\tilde \delta_m ) \  J_{-(m+\alpha)}(kr) \Bigr].
\label{singular}
\end{eqnarray}
Imposing regularity for $f_m(r)$
as $r\to 0$ requires the  phase shift (\ref{phaseshift}) to be
$\delta_m = -(\pi\alpha/2) {\rm sgn}(m+\alpha)$.
Correspondingly, for $R \to 0$,
the scattering amplitude (\ref{central}) is given by
\begin{eqnarray}\label{abscat1}
F(\phi)  &=& \frac{-i}{\sqrt{2 \pi k}} \Biggl[
\left( e^{-i\pi \alpha} - 1 \right) 
\sum_{m = -[\alpha]}^{\infty} e^{i m \phi}
\\ \nonumber &+& \left( e^{i \pi\alpha} - 1 \right) 
\sum_{m = -\infty}^{-[\alpha]-1} e^{i m \phi} \Biggr] ,
\end{eqnarray}
 where $\alpha=[\alpha]+\{\alpha\}$, with integer part $[\alpha]$
and non-integer part $0\le \{ \alpha \} <1$.
Summation of the series in Eq.~\eqref{abscat1} yields\cite{ruijsenaars}
\begin{eqnarray}\label{abf}
F(\phi) &=& \frac{-i}{ \sqrt{2 \pi  k}}
\Bigl ( 2 \pi \delta(\phi) [\cos(\pi\alpha)  - 1 ] \\
\nonumber &+& e^{-i ([\alpha]+1/2) \phi} 
\frac{\sin( \pi\alpha) }{ \sin(\phi/2)} \Bigr) .
\end{eqnarray}
Up to the forward scattering ($\phi = 0$) amplitude,
Eq.~\eqref{abf} reproduces the AB result,\cite{ab,flux,hagen1}
here obtained in terms of scattering phase shifts.
Note that the forward scattering $\delta$-term, 
missing in the AB calculation,\cite{ab}
 naturally appears in our phase shift analysis and 
is essential for establishing unitarity of the scattering
 matrix.\cite{ruijsenaars,sakoda}

In alternative approaches to obtain $F(\phi)$ for 
the ideal solenoid,  following the original AB method,\cite{ab}
the asymptotics of the exact wavefunction is computed 
from its integral representation.
As a result, the incident wave corresponding to Eq.~\eqref{incoming}
has an additional phase factor
$e^{-i\pi \alpha\ {\rm sgn}(\sin\phi)} e^{-i\alpha\phi}$,
i.e., one has a multi-valued incoming plane wave.
The precise relation between these two approaches has been
discussed in several works and is still under
 debate,\cite{ruijsenaars,hagen1,hagen2,hagen3,giacconi,sakoda}
albeit the difference is of little relevance to experimentally observable
quantities.  In particular, the transport cross section 
$\sigma_{\rm tr}$ in Eq.~\eqref{cross} does not depend on the
forward scattering amplitude at all.  We conclude that our approach is
able to reproduce the AB effect,  $\sigma_{\rm tr}
=(2/k) \sin^2(\pi\alpha)$, with oscillations as 
function of the dimensionless flux parameter $\alpha$. 
In particular, $\sigma_{\rm tr}=0$ for integer $\alpha$.

\subsection{Magnetic ring of finite width}\label{sec5c}

Before discussing concrete results for the 
scattering amplitude and the transport cross section in the presence of 
a ring-shaped magnetic field, we 
now generalize the setup to a finite width,
with $R_1<R_2$ denoting the inner and outer radii of the ring, cf.~the
inset of Fig.~\ref{fig2}.
Again, $\Lambda_\phi(r)$ in Eq.~\eqref{lll} is expressed in terms
of a dimensionless flux function $\varphi(r)$.
When $\Lambda_\phi$ is a vector potential, the associated
 magnetic field $B_z(r)=B$ is taken uniform within the ring region
and zero outside; for concreteness, we take $B\ge 0$.
This profile allows for an exact solution,
while more general smooth field profiles
can be treated within the Wentzel-Kramers-Brillouin (WKB) approximation,
see Sec.~\ref{sec5d}.  

We use dimensionless coordinates ($\rho=kr$ and 
${\cal R}_{1,2}=k R_{1,2}$) and flux parameters, 
\begin{equation}
\nu_{1,2} = \frac{\pi B R_{1,2}^2}{\Phi_0} , 
\quad \nu\equiv \frac{\nu_1}{{\cal R}_1^2} = \frac{\nu_2}{{\cal R}_2^2} ,
 \quad \alpha = \nu_2-\nu_1.
\end{equation}
The function $\varphi$ then reads with $r=\rho/k$:
\begin{equation}\label{phi2}
\varphi(r) =  \left \{ \begin{array}{cc} 0, & \rho<{\cal R}_1, \\  
\nu \rho^2 - \nu_1, & {\cal R}_1 < \rho < {\cal R}_2,\\
\alpha , & \rho>{\cal R}_2. \end{array} \right.
\end{equation}
For $R_1\to R_2$, this reduces to Eq.~\eqref{aphi}.
In particular, $\alpha$ in Eq.~\eqref{phi2} 
again denotes the total dimensionless flux.

For given $j = m+1/2$, the components of the Dirac spinor $\psi_m$ obey
Eqs.~(\ref{fmeq}) and (\ref{gmeq}), with Eq.~\eqref{phi2}
now determining $W_m(\rho)$. The solutions for $r <R_1$ and $r >R_2$ are as 
in Eq.~\eqref{fmthin},
\begin{equation} \label{fmeqthick}
f_m (\rho) = \left\{ \begin{array}{cc} a_m  J_m(\rho) , &  \rho < {\cal R}_1, \\ 
b_m  \left( e^{2 i \tilde \delta_m} H_{m+\alpha}^{(1)}(\rho) + 
H_{m+\alpha}^{(2)}(\rho) \right), &\rho > {\cal R}_2, \end{array} \right. 
\end{equation}
where 
$a_m$ and $\tilde \delta_m$ are to be determined, and $b_m$  is
given in Eq.~\eqref{bm}. 
For $R_1 < r < R_2$, Eq.~(\ref{fmeq}) can be 
solved in terms of the  confluent hypergeometric functions
$\Phi$ and $\Psi$,\cite{gradsteyn}
\begin{eqnarray} \nonumber
f_m(\rho) &=& \rho^{|\tilde m|}  e^{- \nu \rho^2 /2} \Bigl[
c_m \Phi(\xi_m, 1+ |\tilde m| ; \nu \rho^2)  \\
\label{inside}
& +& d_m   \Psi(\xi_m, 1+|\tilde m|; \nu \rho^2)
\Bigr] , \\ \nonumber
\xi_m &\equiv& 1 + \tilde m \Theta(\tilde m) -1/4\nu,
\quad \tilde m\equiv m-\nu_1.
\end{eqnarray}
The coefficients $c_m$ and $d_m$, together with $a_m$ and
the phase shift $\tilde \delta_m$ in Eq.~(\ref{fmeqthick}),
follow by matching  $\psi_m$ at $r = R_{i = 1,2}$.
Taking into account that $W_m$ is a continuous function of $\rho$,
we have
\begin{equation}
f_m \left({\cal R}_i^+\right) = f_m \left({\cal R}_i^-\right) ,
\quad f_m'\left ({\cal R}_i^+\right) = f_m'\left ({\cal R}_i^-\right) ,
\end{equation}
where the second condition follows by continuity of the lower 
spinor component $g_m$.
It is convenient to introduce the transfer matrix $\hat {\cal T}_m$
connecting the solutions at 
$\rho = {\cal R}_1^+$ and ${\cal R}_2^-$, 
\begin{equation}
\left( \begin{array}{c} f_m({\cal R}_2^-) \\ f_m'({\cal R}_2^-) 
\end{array} \right) = \hat {\cal T}_m \, 
\left( \begin{array}{c} f_m({\cal R}_1^+) \\ f_m'({\cal R}_1^+) 
\end{array} \right)
= a_m \hat {\cal T}_m \, \left( \begin{array}{c} J_m({\cal R}_1) \\ 
J_m'({\cal R}_1) \end{array} \right).
\end{equation}
Explicitly, the transfer matrix for the magnetic ring of finite width is
\begin{equation}
\hat {\cal T}_m = \left( \begin{array}{cc}
\Phi_2 & \Psi_2 \\ \Phi_2' & \Psi_2'
\end{array} \right) \left( \begin{array}{cc}
\Phi_1 & \Psi_1 \\ \Phi_1' & \Psi_1'
\end{array} \right)^{-1} ,
\end{equation}
where we use the abbreviation
\[
\Phi_{i=1,2} \equiv {\cal R}_i^{|\tilde m|}
 e^{-\nu_i/2} \Phi(\xi_m, 1+|\tilde m| ; \nu_i),
\]
and similarly for $\Psi_i$. We mention in passing that
for the infinitesimally thin magnetic ring in Sec.~\ref{sec5a} 
(where $R_{1} = R_2=R$), 
the transfer matrix is $\hat {\cal T}_m = \left( \begin{array}{cc}
1 & 0 \\ \alpha/{\cal R} & 1 \end{array} \right)$.

For the finite-width ring,
using Eq.~(\ref{fmeqthick}) the phase shift $\tilde \delta_m$ is then again 
given by Eq.~(\ref{dmthin}), with ${\cal R} \to {\cal R}_2$ 
and the logarithmic derivative ${\cal L}_m$ replaced by
\begin{equation} \label{calL2}
{\cal L}_m = \frac{u_{m,2}}{u_{m,1}} , \quad
\left( \begin{array}{c} u_{m,1}\\ u_{m,2} \end{array}\right)
 = \hat {\cal T}_m \left( \begin{array}{c} J_m({\cal R}_1) \\ J_m'
({\cal R}_1) \end{array} \right) .
\end{equation}

With the above expressions, it is straightforward to compute the
scattering phases $\delta_m=\tilde \delta_m-\pi\alpha/2$ numerically
for the finite-width ring geometry. Thereby
we obtain\cite{foot} the scattering amplitude $F(\phi)$ from 
Eq.~\eqref{central} and the transport cross
 section $\sigma_{\rm tr}$ from Eq.~\eqref{cross1}. 

\begin{figure}
\includegraphics[width=0.45\textwidth]{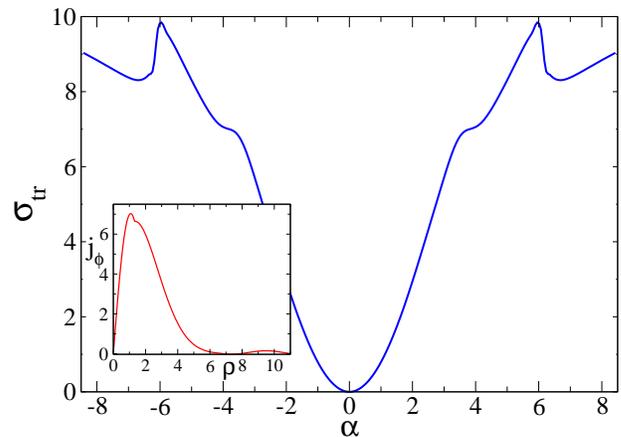}
\caption{\label{fig3} (Color online) The main panel is as in Fig.~\ref{fig2}
but for $kR_1=1.32$ and $R_2=5R_1$.  
Inset: Current density $j_\phi(\rho)=\hat e_\phi\cdot {\bm j}$ 
vs radial coordinate $\rho$ for the quasi-bound state 
with $j=3/2$ present at $\alpha\simeq 6$. } 
\end{figure}

Numerical results obtained under this approach
are shown in Figs.~\ref{fig2} and \ref{fig3}.  First, in the main 
panel of Fig.~\ref{fig2},
we show the transport cross section $\sigma_{\rm tr}$ as
a function of the total flux $\alpha$.  In this example, both radii $R_1$
and $R_2$ were chosen very small, such that scattering by the ring
is close to the one by an ideal AB solenoid.   As a consequence, we observe
the AB oscillations with unit flux period. 
In contrast to the ideal AB result, a complete suppression of scattering
for $\alpha\in \gz$ is observed in the finite-width ring 
only for $\alpha=0$, while the maximum value
$\sigma_{\rm tr}=2/k$ for half-integer $\alpha$ is still perfectly realized.
In fact, the phase shift analysis in Sec.~\ref{sec5b} shows that 
a given oscillation period is  determined by one specific $m$ value in the
ideal AB case.  For the non-ideal finite-width ring, other total
angular momenta also start to contribute, and this mixing effect
destroy the perfect constructive interference needed for $\sigma_{\rm tr}=0$.
On the other hand, the destructive interference responsible for the maxima
of $\sigma_{\rm tr}$ at half-integer $\alpha$ is more robust  since it is
dominated by a single $m$ value.

In Fig.~\ref{fig3}, we study scattering by a much larger ring.
In this case, the AB effect is absent, which can be understood by
noting that the Fermi wavelength ($2\pi/k$) of the particle is now smaller
than the outer circumference $2\pi R_2$ of the ring.  Quantum interference
of waves surrounding the obstacle in opposite directions is then
largely averaged out, and, moreover, the wavefunction
can partially penetrate into the ring area.
However, a remarkable peak feature at $\alpha\approx 6$ appears now
in the transport cross section, see Fig.~\ref{fig3}.  
This feature can be traced to the appearance
of a quasi-bound state with $j=3/2$ at this flux value (for the
considered energy), which then causes a scattering resonance, cf.~our
discussion in Sec.~\ref{sec5d}. The 
inset of Fig.~\ref{fig3} shows the current density profile 
for precisely this quasi-bound state.  While the radial component vanishes,
$j_r=0$, we find a circularly oriented current, $j_\phi \ne 0$, which 
is mainly localized inside the ring ($r<R_1$) 
and represents a current-carrying bound state.  
We note that more quasi-bound states appear for 
larger $\alpha$, causing additional peak features in  
$\sigma_{\rm tr}(\alpha)$ beyond those shown in Fig.~\ref{fig3}.

\subsection{Quasi-bound states and scattering resonances}\label{sec5d}

\begin{figure}
\includegraphics[width=0.45\textwidth]{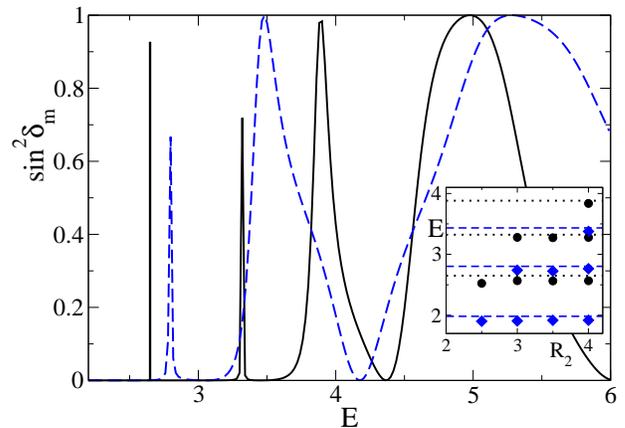}
\caption{\label{fig4} (Color online) 
Partial cross section $\sin^2 \delta_m$
 vs energy $E$ for a ring-shaped confinement as in Sec.~\ref{sec5c}.
The numerical results are for total angular momentum states with
$m=1$ (solid black) and $m=-1$ (dashed blue curve).
The radii are $R_2=7R_1$ and $R_1=0.5\ell_B$ with
$\ell_B= \sqrt{2c/eB}$, and energies are in units of $\hbar v_F / \ell_B$.
Inset: WKB results for quasi-bound state energies $E_r$
 vs $R_2$ (lengths in units of $\ell_B$), for $m=1$ (black circles) 
and $m=-1$ (blue diamonds) with fixed $R_1=0.5\ell_B$.
For comparison, the exact levels for infinite $R_2$ from 
Ref.~\onlinecite{ademarti2} are shown for $m=1$ (dotted black) 
and $m=-1$ (dashed blue curve).
}
\end{figure}

The magnetic confinement built up by the ring-shaped field
can generate quasi-bound states, which for $R_2\to \infty$
become true bound states.\cite{ademarti,ademarti2}
The quasi-bound state spectrum then causes resonances
in the scattering amplitude when the energy $E=\hbar v_F k$ is varied.
For given total angular momentum $j=m+1/2$, 
the corresponding phase shift $\delta_m(E)$ goes
 through the value $\pi/2$ as $E$ crosses a resonance level $E_r$.
The corresponding resonance width $\Gamma_r$ can be estimated from\cite{newton}
\[
(d/dE) \cot \left[ \delta_m(E=E_r)\right] = - 2/\Gamma_r.
\]
To access these resonances, we first put Eq.~(\ref{fmeq}) into a 
canonical form with separated kinetic and potential energy terms.
The substitution $f_m(\rho) = \rho^{-1/2} \tilde f_m(\rho)$ yields
\begin{equation} \label{fmeq2}
\left[ - \partial^2_\rho + V_m(\rho) \right] \tilde f_m = \tilde f_m ,
\quad V_m \equiv W_m^2+W'_m,
\end{equation}
where $V_m(\rho)$ is an effective potential energy
for the radial motion and the lower spinor component is
$g_m = \rho^{-1/2} ( -\partial_\rho + W_m ) \tilde f_m$.
In this form, Eq.~(\ref{fmeq2}) can be treated within the standard 
WKB approach, which represents an attractive 
alternative to semiclassical approaches to the Dirac equation as it 
avoids the appearance of non-Abelian Berry phases.\cite{kory,keppeler,carmier} 
For a  magnetic ring as in Sec.~\ref{sec5c}, the effective potential $V_m$ 
has a hard repulsive core for $r\to 0$ plus a barrier at larger distances, 
i.e., a quantum well is formed with classically allowed motion for 
$r_0<r<r_1$. The ``turning points'' $r_{0,1}$ here depend on the
energy $E=\hbar v_F k$ under consideration.  
For finite $R_2$, this barrier is of finite width
and quasi-bound states within the well region may exist.
The classically forbidden region $r_1<r<r_2$ (where $r_2$ is another
turning point) then corresponds
to tunneling trajectories where the ``particle'' escapes from the well region.
For $R_2\to \infty$, the barrier becomes infinitely wide and this
escape probability vanishes, i.e., we obtain true bound states in the well
region.  Using the radial variable $r=\rho/k$, Eq.~(\ref{fmeq2}) reads
\begin{eqnarray}\label{fmeq3}
&& \left[ - \partial^2_r + U_m(r) \right] \tilde f_m(r) = 
\epsilon \tilde f_m(r) ,\\ \nonumber
&& U_m(r)= w_m^2 + \partial_r w_m, \quad
w_m(r) = k W_m(kr),
\end{eqnarray}
where the modified Bohr-Sommerfeld quantization condition
for the \textit{complex-valued 
``energy''}\ $\epsilon\equiv k^2$ is\cite{mur}
\begin{eqnarray}\label{complex}
&& \int_{r_0}^{r_1}dr\ \sqrt{\epsilon-U_m(r)}  = \pi \left( n + 
\frac12 -\frac {\chi(a) }{2 \pi} \right),\\ \nonumber
&& \ \chi(a)  = \frac{1}{2i} \ln \left(
\frac{\Gamma(ia + 1/2)}{\Gamma(-ia+1/2) 
\left[1 + e^{-2 \pi a} \right ]} \right)\\  \nonumber
&&\qquad \quad + \  a (1 - \ln a), \\ \nonumber
&&  \  a = \frac{1}{\pi} \int_{r_1}^{r_2} dr \ \sqrt{ U_m(r)-\epsilon },
\end{eqnarray}
with $n = 0,1,2,\ldots$ and the Gamma function $\Gamma(z)$.
The complex resonance values for $\epsilon$ 
solving Eq.~\eqref{complex} can be found numerically.
Equation \eqref{complex} is formally exact for the case of a 
parabolic barrier, but also applies for an arbitrary smooth potential
and is expected to remain accurate\cite{mur} even for small $n$.
We now write 
\[
\epsilon = (k - i \gamma/2)^2 \approx k^2 - i k \gamma.
\]
For a quasi-bound level with energy $E_r=\hbar v_F k$,
the resonance width is then $\Gamma_r=\hbar v_F \gamma$.
Using ${\rm Im} \chi (a) \approx e^{-2 \pi a}/2$ for $a\agt 1$, we obtain 
\begin{equation}
\Gamma_r/\hbar = T^{-1}_k e^{-2 \pi a} ,
\end{equation}
where the period of radial motion is
\[
T_k = \frac{ 2k}{v_F} \int_{r_0}^{r_1} \frac{dr}{\sqrt{k^2 - U_m(r)}}.
\]
Our numerical results for the partial cross section, $\sin^2\delta_m$, 
as a function of energy, and the WKB results for
the corresponding quasi-bound state energies $E_r$ 
 are shown in Fig.~\ref{fig4}.  Here we take
the field profile as in Sec.~\ref{sec5c}.\cite{foot2} 
With increasing $R_2$ (keeping $R_1$ fixed), new quasi-bound energy
levels localized in the well region appear, see inset of Fig.~\ref{fig4}.
Very good agreement with exact quantum calculations\cite{ademarti2} 
for the infinite barrier case ($R_2\to \infty$) is observed, i.e.,
these energy levels remain basically unchanged when increasing $R_2$.
The only noticeable deviation from the exact spectrum of 
Ref.~\onlinecite{ademarti} is seen for $m=0$, where the potential 
$U_{m=0}(r)$ creates an infinitely attractive well for $r\to 0$.
In that case, the WKB approximation becomes
questionable in that ``steep'' region.
The main panel in Fig.~\ref{fig4} illustrates the sequence of quasi-bound
states present because of the magnetic confinement.  The corresponding
scattering resonances appear as peaks in the transport cross section
$\sigma_{\rm tr}$ when varying energy or the effective flux parameter 
$\alpha$.

\section{Concluding remarks} \label{sec6}

In this paper, we have studied scattering of massless two-dimensional Dirac
fermions by magnetic perturbations of various types. The
model is applicable to quantum transport in monolayer graphene
and for the surface state of strong topological insulators.
The magnetic fields can correspond to orbital or Zeeman fields,
strain-induced fields in graphene, or exchange fields generated
by ferromagnets.

The full scattering solution was discussed in detail for radially
symmetric perturbations, where the scattering amplitude can be
expressed in terms of phase shifts in a given total angular 
momentum channel,  and within the Born approximation for
the general case.  Our approach now allows for a systematic study
of the scattering of Dirac fermions on magnetostatic perturbations. 

As applications, we have studied scattering
by magnetic dipoles within the Born approximation, and 
fully nonperturbative scattering for the case of ring-shaped magnetic fields.
The Born approximation is only valid when the perturbation 
has zero total flux ($\alpha=0$). For the magnetic dipole, we have pointed out
characteristic angular dependencies in the differential cross section
 that may allow to unambiguously identify
massless Dirac fermions.  For the ring-shaped field case, as one increases
the lateral size ($R_2$) of the magnetic perturbation, we have a crossover
from the Aharonov-Bohm case to a regime dominated by scattering resonances.
In the first case, $R_2\to 0$, particle trajectories surround the
flux region but essentially do not penetrate it, leading to 
the oscillatory transport cross section $\sigma_{\rm tr}\propto 
\sin^2(\pi \alpha)$. In the second case, where the particle wavelength is small
against the size of the perturbation, $kR_2>1$, the AB oscillations in 
$\sigma_{\rm tr}(\alpha)$ are absent. However, now quasi-bound states arise 
due to the magnetic confinement, causing scattering resonances which show
 up as peaks in $\sigma_{\rm tr}(\alpha)$.

To conclude, we hope that these predictions motivate further theoretical
work and that they will be tested experimentally 
in the near future.

\acknowledgments
We thank A. De Martino for discussions
and acknowledge financial support by the DFG Schwerpunktprogramm 1459.


\begin{thebibliography}{10}
\bibitem{foot} In the numerical evaluation, we sum over all 
angular momentum states $|j|<40$ (with $j=m+1/2$).  Scattering phases
with very large $|j|$ are difficult to compute reliably yet average
out in practice.
\bibitem{foot2} In order to avoid artefacts caused by treating sharp
boundaries within the WKB approach, we have used a smoothening of the
field steps.
\end{thebibliography}
\end{document}